\documentstyle[11pt,epsf]{article}
\def\beq{\begin{eqnarray}}
\def\eeq{\end{eqnarray}}
\def\bea{\begin{eqnarray*}}
\def\eea{\end{eqnarray*}}

\def\centeron#1#2{{\setbox0=\hbox{#1}\setbox1=\hbox{#2}\ifdim
\wd1>\wd0\kern.5\wd1\kern-.5\wd0\fi
\copy0\kern-.5\wd0\kern-.5\wd1\copy1\ifdim\wd0>\wd1
\kern.5\wd0\kern-.5\wd1\fi}}
\def\ltap{\;\centeron{\raise.35ex\hbox{$<$}}{\lower.65ex\hbox{$\sim$}}\;}
\def\gtap{\;\centeron{\raise.35ex\hbox{$>$}}{\lower.65ex\hbox{$\sim$}}\;}

\def\singleandthirdspaced{\baselineskip=\normalbaselineskip\multiply
    \baselineskip by 130\divide\baselineskip by 100}
\def\singlespaced{\baselineskip=\normalbaselineskip}

\newcommand{\newc}{\newcommand}
\newc{\qbar}{{\overline q}}
\newc{\Kahler}{K\"ahler }
\newc{\deltaGS}{\delta_{\rm GS}}
\begin{document}
\begin{titlepage}
\begin{flushright}
{\large hep-th/0405190 \\ SCIPP-2004/16\\
}
\end{flushright}

\vskip 1.2cm

\begin{center}

{\LARGE\bf Catastrophic Decays of Compactified Space-Times\\ }

\vskip 1.4cm

{\large  Michael Dine, Patrick J. Fox and Elie Gorbatov}
\\
\vskip 0.4cm
{\it Santa Cruz Institute for Particle Physics,
     Santa Cruz CA 95064  } \\

\vskip 4pt

\vskip 1.5cm

\begin{abstract}
Witten long ago pointed out that the simplest Kaluza-Klein theory,
without supersymmetry,
is subject to a catastrophic instability.  There are a variety of
string theories which are potentially subject to these
instabilities.  Here we explore a number of questions:  how generic
are these instabilities?  what happens when a potential is generated
on the moduli space?  in the presence of supersymmetry breaking,
is there still a distinction between supersymmetric and non-supersymmetric
states?
\end{abstract}

\end{center}

\vskip 1.0 cm

\end{titlepage}
\setcounter{footnote}{0} \setcounter{page}{2}
\setcounter{section}{0} \setcounter{subsection}{0}
\setcounter{subsubsection}{0}

\singleandthirdspaced

\section{Introduction}

In studies of string theory in four dimensions, we often focus on
string vacua which, at the classical level, have $N=1$ supersymmetry.
But it has never been clear why these should be particularly important.
It is true that $N=0$ vacua have potentials on moduli space at one loop,
but the supersymmetric vacua have potentials non-perturbatively.
The non-supersymmetric vacua
often have tachyons in their moduli spaces, but it seems plausible
that these theories might have AdS minima, somewhere on the interior of
moduli space, and perhaps local dS or Poincare invariant minima elsewhere.
In fact, even distinguishing these vacua seems to require some
awkward phraseology, such as theories with ``approximate moduli spaces
with approximate $N=1$ supersymmetry."

But various authors\cite{wittenkk,fh,harvey,brillhorowitz,dealwis} have noted one
possible, sharp distinction: non-supersymmetric vacua sometimes suffer from a
bizarre and catastrophic non-perturbative instability, known as Witten's
``bubble of nothing" (BON)\cite{wittenkk}.  In these
vacua, a bubble forms, just as in conventional false vacuum decay.
This bubble grows at the speed of light, but the bubble wall,
from the perspective of a four dimensional observer,
is the end of space-time.  The authors of refs.
\cite{fh,harvey} even speculated that this might be generic to
non-supersymmetric states.  In \cite{dealwis},
further examples were provided, and a connection between
theories with BON's and closed string tachyons
was conjectured.  Such bubbles do not seem to arise for
supersymmetric solutions.  It is natural to speculate that this represents a
real distinction between supersymmetric and non-supersymmetric states.  One
might go so far as to suggest that the non-supersymmetric states do not make
sense.  Indeed, in the particular case of toroidal compactification, this was
the viewpoint espoused by Brill and Horowitz\cite{brillhorowitz} when they
extended Witten's analysis to construct solutions with arbitrarily negative
energy.  If the existence of such bounce solutions (and their associated
negative energy configurations) is more general, it is even tempting to argue
that this might demonstrate that string theory {\it predicts} what we usually
call low energy supersymmetry.

It is the purpose of this note to pursue these observations and speculations
further.  There are a number of objections which can (and have) been raised
to these sorts of ideas.  All of these must be addressed if one is to
argue that the Witten bubble is a significant pathology which
indicates that broad classes of string vacua are inconsistent -- and perhaps
to make a prediction of low energy supersymmetry:
\begin{enumerate}
\item  Is the Witten solution, discovered by examining
the lowest order classical equations of Einstein's gravity, a
solution of string theory?
\item  Is the Witten solution generic to
non-supersymmetric solutions of string theory?  At present, it can only be
written down in limited classes of string/M theory models.
\item  At the quantum level, the moduli are not
exact, already at one loop.  As a result, motion on the moduli space occurs on
a much more rapid time scale than vacuum decay.  So
in what sense is it meaningful to discuss the Witten bounce in a quantum
theory of gravity\cite{banks}?  
\item If supersymmetry is spontaneously broken (by a small amount, as we
  imagine might be relevant to the description of the real world) does a
  Witten instability appear?  In other words, is there any real distinction
  between classically supersymmetric vacua with four supersymmetries and
  non-supersymmetric vacua?
  \item  Finally, like other gravitational instantons, there is no sharp argument
  that the BON must be included in the semiclassical approximation to some
  underlying quantum theory of gravity.  Perhaps it can simply be excluded from consideration,
  as can the negative energy configurations.
\end{enumerate}

The goal of this work is to address these questions.  In most cases, we will
not be able to give definitive answers, but we will argue that it is quite
plausible that the Witten solution is a solution of string theory, that it is
exists in a broad class of non-supersymmetric strings, that it is relevant
even in the presence of a potential for the moduli, and that it will not
appear after (small) spontaneous supersymmetry breaking.

In the next section, we review the original Witten solution, and in
section 3 we describe it in four dimensional terms, along the lines of
the analysis of Coleman and DeLuccia (CDL)\cite{cdl}.  In section 4
we discuss the BON as a solution of string theory, arguing that, at
least to all orders in the $\alpha^\prime$ expansion, it is solves the
classical equations, and that, in the non-supersymmetric case, there
is a sensible, modular invariant conformal field theory which
describes these configurations.  We then consider the question of
whether such instabilities are generic.  Our starting point is the
work of \cite{fh}, in which non-supersymmetric configurations of the
eleven-dimensional heterotic string were shown to admit BON solutions.
We explain why it is difficult to prove that similar instabilities
exist in general weakly coupled string theories -- but why it seems
likely that they do.  We then turn to the question: what happens when
moduli are stabilized.  We start with the following observation: {\it
if} some non-supersymmetric string theory solution describes nature,
then necessarily there is a stable minimum for the moduli.\footnote{Non-supersymmetric
string states often have tachyons in regions of the moduli space, and naively
one might then expect that if sensible quantum theories exist,
the moduli will be stabilized with negative cosmological constant.
This point has been stressed in \cite{dealwis}.}    We write
down a model potential, which has stable moduli and the expected
asymptotic behavior, and proceed to construct BON solutions, paying
particular attention to the required boundary conditions.  We find
that such solutions generally exist, not only when the four
dimensional space is flat, but also if it is de Sitter or Anti-de
Sitter.  We note that this semiclassical analysis is not likely to be
reliable, but argue that it strongly indicates that the instability
will be present in any such case.  Finally, we explain why vacua with
approximate supersymmetry (of the sort one expects if there is some
sort of low energy supersymmetry) are not subject to these
instabilities.

In the final section, we discuss possible ways to extend this analysis.  Particularly
interesting is the question of such bounce solutions
in flux vacua\cite{fluxvacua,kklt}.  We also
enumerate some puzzles raised by this work.  In all of these
cases, if we assert that the Witten solution does represent an instability,
we have to ask how one might have
gotten into the ``false vacuum" in the first
place.  We argue that one possible way to understand
this is in terms of the time-symmetric extensions of these
solutions studied in \cite{evaetal}.  Such solutions, however,
are also subject to such instabilities, and their
ultimate fate is not clear.  Related issues include
questions concerning the
AdS-CFT correspondence.  Finally, we speculate on the implications of the
bubble of nothing for string phenomenology.

\section{The Witten Bounce}

Coleman and Deluccia\cite{cdl}, many years ago, formulated the
problem of vacuum decay in theories of gravity.  They focused on
theories in four dimensions, and considered decays of states with
positive, zero and negative cosmological constants.  Gravity, they
discovered,
introduces dramatic effects.  For example, they found that often
decay from flat space or de Sitter space to
anti-de Sitter space does not occur, and when it does it leads to
catastrophic gravitational collapse.  Banks has recently revisited
many of these issues, raising very general questions about vacuum
decay in general relativity\cite{banks}.

Shortly afterwards, Witten wrote down a bounce solution of a different
sort\cite{wittenkk}.  He noted that the Euclidean Schwarschild solution
of five
dimensional gravity is a bounce, which describes the decay of the Kaluza-Klein
vacuum, the compactification of five dimensional
gravity on a circle of radius $R$.  This solution is given by
\beq\label{eqn:wittenmetric}
ds^2 = {dr^2 \over (1-{R^2 \over r^2})}+ r^2 d\Omega_3^2 + (1-{R^2 \over
r^2})dy^2.
\eeq
Here $y$ is the coordinate of the fifth dimension, $0<y <2\pi R$
and $d\Omega$ is the element of area of $S^3$.
As we will discuss in greater
detail shortly, from a four-dimensional perspective, the
coefficient of $dy^2$ is a field, the radial modulus.
This configuration has finite action,
\beq
S_o={\pi R^2 \over 4 G}.
\eeq   
The singularity at $r=R$ is a
coordinate singularity; in the neighborhood of $r=R$,
the metric can be written in the form
\beq
ds^2 = du^2 + u^2 d\phi^2+ R^2 d\Omega_3^2,
\eeq
with $y = R\phi$,
i.e. the space looks like $R^2 \times S^3$.

Witten showed that this solution describes vacuum decay.  To
see the structure of the bubble which forms,
one passes to Minkowski signature
by taking
polar angle of $S^3$, $\theta$, and continuing:
\beq
\theta \rightarrow i\psi + {\pi \over 2}.\eeq
Then the metric becomes:
\beq
ds^2=-r^2 d\psi^2 + r^2 \sinh^2(\psi)d\Omega^2 + {dr^2 \over (1-{R^2 \over
r^2})} + (1-{R^2 \over r^2}) dy^2.
\eeq
This describes a ``bubble of nothing" (BON).  It is a non-singular,
geodesically complete space.  $r=R$ is not a singular surface--
it is, from the perspective of a four dimensional observer,
the end of space-time.

Witten observed another striking feature of this solution: it is
topology changing and it is not admissible in supersymmetric theories.
To see this, note that if one does a $2\pi$ rotation near
$r=R$, fermions pick up a minus sign.  Since this region is smoothly
connected to the large $r$ region, the fermions in the original
Kaluza-Klein space must obey anti-periodic boundary conditions in
$\phi$.

While a BON is a troubling end to a universe,
it has never been clear whether one can simply discard theories
with such instabilities.  First, the actual fate of the universe
is obscure.  If such bubbles make sense, then they will be nucleated
with a certain probability per unit volume per unit time, and the expanding
bubbles will collide.  Witten speculates that this situation will
result in gravitational collapse (a two-bubble solution has been
exhibited recently by Horowitz and Maeda\cite{hm}).  Second,
one might conjecture
that, as we expect that string theory resolves at least some cosmological
singularities, it might resolve these.  If this is the case,
then perhaps we live in such a universe, and its lifetime is just extremely
long. 

However, Witten pointed out that the existence of these solutions is related
to another seeming pathology of the Kaluza-Klein space:  the energy
is unbounded below.  Explicit configurations of negative energy were
exhibited in \cite{brillhorowitz}.  
\beq
ds^2 = U d\chi^2 + U^{-1}dr^2 + r^2 d\Omega^2
\eeq
where 
\beq
U(4r)= 1 - {2m \over r} - {q^2 \over r^2}
\eeq 
The energy here is just $m$, which can have either sign.
Not surprisingly, the zero time ($\tau =0$)
configuration in Witten's solution corresponds to the zero energy
configuration (with $q=R$).

A second issue which has obscured the significance of the BON has to do
with moduli.  From the perspective a four dimensional observer, the
bubble involves excitations of the graviton and a scalar, the radial
modulus.  In string theory, such a modulus will acquire a potential
at one loop, and this will lead to motion on a scale rapid compared to
the decay time.  On the other hand, if such configurations have something
to do with the world we see around us, there must be a stable minimum
for the moduli potential.  To analyze this problem, as well as to consider
possible other solutions, it will be helpful to understand the Witten bubble
from a four dimensional perspective.

\section{The Witten Bubble From a Four Dimensional Perspective}

One can try to formulate the Witten solution directly in the effective four
dimensional field theory.  Reducing from five dimensions to four, one has the
four dimensional graviton, a scalar field, and a gauge field.  In the
solutions we will consider, the gauge field is not excited, and we will ignore
it.  We parameterize the five dimensional metric as
\beq
g_{55} = R^2(x) =Re^{{2 \over \sqrt{3}} \phi(x)}
\eeq
(here $\phi$ is a four dimensional field, not to be confused with the five
dimensional angle; similarly, $R(x)$ is a field; we will sometimes use the 
letter $R$ to refer to this field and sometimes to the
value of the radius in the would-be vacuum).  The four dimensional effective action, after Weyl
rescaling, has the form
\beq
S = \int d^4 x \sqrt{g}(R + {1 \over 2}\partial_\mu \phi \partial_\phi
\phi g^{\mu \nu}).
\eeq

In this form, we have a four-dimensional effective action of the type
considered by
CDL.  Their action also includes a potential, $V(\phi)$.
CDL look for a solution with $O(4)$ symmetry, which interpolates
between the false and true vacua.  The symmetry dictates that the
metric may be written in the form:
\beq
ds^2 = d\xi^2 + \rho^2(\xi)  d\Omega_3^2.
\label{cd2}
\eeq
The Euclidean equations for $\phi$ and $\rho$ are:
\beq
\phi^{\prime \prime} + {3 \rho^\prime \over \rho} \phi^\prime =
{dV \over d \phi},
\label{eqn:cd3}
\eeq
and
\beq
\rho^{\prime ~2} = 1 + {1 \over 3} \kappa \rho^2 ({1 \over
2}\phi^{\prime~2}-V).
\label{eqn:cd4}
\eeq

CDL then consider an analog problem, where $\xi$ plays the role of
time.  They require that at $\xi=0$, the system starts off in
(near) the true vacuum, while for $\xi\rightarrow \infty$ the
system tends to the false vacuum.  They also require
that the system start off with zero velocity.

\begin{figure}
\centerline{\epsfxsize=4 in \epsfbox{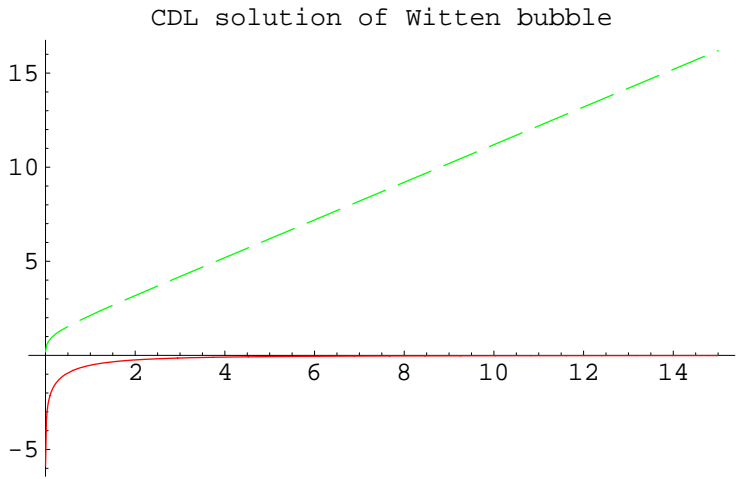}}
\caption{The Witten bubble as a CDL problem: the dotted line is $\rho(\xi)$
  and the solid line is $\phi(\xi)$.}\label{fig:flatbubble}
\end{figure}

How does the Witten solution look in this description? 
For the Witten problem, the potential vanishes and
the ``true vacuum" lies infinitely far away (at $\phi= -\infty$).  Weyl rescaling
to four dimensions, the metric becomes:
\beq
ds^2 = {dr^2 \over \sqrt{1-R^2/r^2}} + {r^2  \sqrt{1-R^2/r^2}
}d\Omega^2.
\eeq
To relate this to the CD equations, we need to solve
\beq
{d \xi \over dr} = (1-R^2/r^2)^{-1/4}.
\eeq
Then we can read off $\rho$ from the metric.  In particular, as $r
\rightarrow R$, we have
\beq
r =R +  \left({3 \over 4} \left(\frac{2}{R}\right)^{1/4} \xi\right)^{4/3}
\eeq
and
\beq
\rho = r\left(\frac{2(r-R)}{R}\right)^{1/4}~~~~~~\phi = \frac{1}{2} \sqrt{\frac{3}{2}} \ln2 \frac{r-R}{R}.
\eeq
Note, in particular, that as $\xi \rightarrow 0$, the field and
its derivative become large.  In the analog mechanics problem,
this corresponds to starting off with a big kick.  It is a simple
exercise with Mathematica to find the Witten solution from these equations,
taking the limiting behavior above for the solution.
Figure~\ref{fig:flatbubble} shows the solution for $\phi$ and $\rho$ as a
function of $\xi$, the CDL coordinate.
The continuation to Minkowski signature is done just as in the five dimensional
case.  One takes the polar angle of $S_3$, and again makes the continuation
$\theta \rightarrow i{\pi \over 2} + \psi$.

Written in this way, there is no particular obstacle to looking
for BON solutions in the presence of a potential.   It will be
necessary, however, to pay close attention to the boundary
conditions.  We will do this shortly.
\subsection{Geodesic motion}

In order to understand the motion of the bubble wall, and light and
matter in the presence of the bubble we need to know the geodesics for
the spacetime.  In the case of the Witten Bubble in flat space these
have been calculated in \cite{Brill:rp}.  As a warm up we rederive
their result, using a method that is easily generalised to
the AdS case.  First, we assume, without loss of generality, that we
fix one of the angles on the sphere, $\theta_1=\pi/2$.  We can then
rewrite Eq.~(\ref{eqn:wittenmetric}) in a more convenient form using
the following coordinate tranformations,
\beq\label{eqn:transforms}
du=\sinh\psi d\theta -d \psi,\ \  dv=\sinh\psi d\theta +d\psi,\ \ z^2=r^2-R^2.
\eeq
The metric is now,
\beq\label{eqn:metricprime}
ds^2=(z^2 +R^2) du dv + dz^2 +\frac{R^2z^2}{z^2+R^2}dy^2
\eeq
There are a set of conserved momenta associated with geodesic motion,
\beq
(z^2+R^2)\dot{u}= k_v,\ \ (z^2+R^2)\dot{v}=k_u,\ \
\frac{R^2 z^2}{z^2+R^2}\dot{y}=k_y
\eeq
where dot corresponds to derivative with respect to the proper time
along the geodesics.  We normalize this affine parameter along the
geodesics such that $\mu^2\equiv\dot{x}_\alpha\dot{x}^\alpha$ is $0\ 
(-1)$ for a lightlike (timelike) geodesic.  Using the conserved
quantities it is possible to write down a first order differential
equation for the radial motion,
\beq
\dot{z}^2+\frac{k_u k_v}{z^2+R^2}+ k_y^2\frac{z^2+R^2}{R^2 z^2}=\mu^2=\left\{
\begin{array}{r@{\quad:\quad}l} 0 & \mathrm{null}\\
-1 & \mathrm{timelike}\end{array}\right.
\eeq
This equation describes a particle moving in an effective potential, 
\beq
V_{eff}=\frac{k_u k_v}{z^2+R^2}+ k_y^2\frac{z^2+R^2}{R^2 z^2}-\mu^2
\eeq

For simplicity we will consider motion where the remaining angle, $\theta$, is
held fixed.  In this case $k_u=-k_v$ and the first term in the effective
potential is negative asymptoting to zero at large $z$, the second term is
positive definite and tends to a constant, $k_y^2/R^2$, at large $z$.  The
existence of unbound motions then depends only on whether $k_y=0$ or not.  For
$k_y=0$ there exist unbound lightlike geodesics but regardless of the
magnitude of $k_y$ there are no unbound timelike geodesics.  

A massive particle carries out oscillatory motion with turning points found by
solving $V_{eff}=0$.  In the case of no motion in the $y$ direction, $k_y=0$,
the inner turning point\footnote{Note that the transformation of
  Eq.~(\ref{eqn:transforms}) is only valid for $r=R$ if $\dot{r}=0$, so if the
  particle makes it into $z=0$ it must stop there.} is $z=0$, the bubble wall.
The bubble wall hits a massive particle which bounces off and moves ahead of
the bubble.  The wall accelerates and catches up, pushing the particle forward
again.  The maximum separation of the particle from the bubble wall is given by
the larger root of $V_{eff}=0$.  With $k_y\neq 0$ the particle has momentum in
the compact extra dimension and the turning points are shifted.  The inner
turning point is shifted out from $z=0$ and the outer is shifted in.

\section{Bubbles of Nothing in String/M Theory}

\subsection{The Five Dimensional Solution as a Solution of String Theory}

Witten's solution would seem to be a good solution of certain
string theories.  For example, Rohm long ago wrote down string
solutions of the Type II and heterotic theories with
Scherk-Schwarz boundary conditions\cite{rohm}.  These have antiperiodic
boundary conditions for the fermions in the fifth dimension,
precisely what would seem to be required to support a Witten
bubble.  Higher order terms in the curvature
will appear in the $\alpha^\prime$ expansion,
and will correct the solution.
But there is no apparent obstacle to constructing a solution in powers of
$1/R$ in the $\alpha^\prime$ expansion.  There are no dangerous zero
modes.  The solution is non-singular, as we saw above.

It is easy to
check, in fact, that all curvature invariants are non-singular for
the Witten solution.
Certain components of the Riemann tensor for Witten's solution blow up or
vanish as $r \rightarrow R$.  For example:
\beq
R_{r \theta r \theta}= -{R^2 \over r^2-R^2}~~~~~R_{\theta \phi \theta \phi}=
-\frac{(r^2-R^2)R^2}{r^4}
\eeq 
So one might worry that terms in the action constructed out of $R_{MNOP}$ might
be singular.  But, for instance, powers of $R_{r \theta r \theta}^2$
are accompanied by $(g^{rr})^2 (g^{\theta \theta})^2$, and thus are
not singular as $r \rightarrow R$.  Instead the scalar quantitity
built from $R_{r\theta r\theta}$ is $-1/R^4$ in the limit
$r\rightarrow R$.  Similarly, to build an invariant from the vanishing
quantity $R_{\theta \phi \theta \phi}$ introduces powers, $(g^{\phi
\phi})^2$, which is singular, but again the resultant invariant is
$-1/R^4$.  From the perspective of a four dimensional observer, this
is rather miraculous.  The four dimensional action involves
combinations of the curvature and the moduli in just such a way that
there are no singularities, no matter how many powers of fields and
derivatives are included.  From the perspective of the five
dimensional observer, there is, of course, no miracle at all; as we
saw, the singularity is a coordinate singularity.

We should also ask whether these are solutions non-perturbatively in
$\alpha^\prime$.  Here we would note, again, that there are not candidate zero
modes which could serve as an obstruction to finding
a solution.  Still, it
would be desirable to have another argument directly in conformal field
theory.

Finally, we should worry about the dilaton and other moduli (besides the 
five dimensional radius).  If we compactify, for example, a ten dimensional
heterotic string on a six dimensional torus, where the torus has the structure
$S^1 \times T^5$, the moduli space factorizes.  The
non-trivial field configuration for $g_{55}$, the radius of the circle,
does not serve as a source for either the dilaton or the other moduli.

\subsection{Generalizations of the Witten Solution}

It is natural to wonder whether one can write such solutions for more general
non-supersymmetric strings.  In weakly coupled string
theories, there are a large number of rather trivial generalizations,
arising whenever one compact dimension is a circle with
anti-periodic boundary conditions for fermions.
The discussion of the previous section suggests another set
of generalizations in weakly
coupled string theory.  We have written Witten's solution purely in terms
of four dimensional fields.  The action for the string dilaton is (apart
from a numerical factor) identical to that for the field $R$.  So we can
simply transcribe our solution in terms of the four dimensional metric
and $R$ into a solution of the four dimensional metric and the dilaton.
In this way, we can write down a host of possible solutions.  For example,
we might consider the $O(16) \times O(16)$ theory compactified on a Calabi-Yau
space (such a configuration is a solution of the classical string equations;
some of its features will be discussed elsewhere\cite{kramer}).
All of these would then admit a solution of the classical equations,
at the level of the two derivative terms.

The difficulty with these solutions is that the four dimensional curvature
is singular, and we have, a priori, no reason to expect a cancellation
of all singular terms in the action by powers of the dilaton,
as occurred in the five dimensional compactifications.

On the other hand, we know that at strong coupling, the dilaton is often the
radius of a circle.  Indeed, Fabinger and Horava generated a large class of
additional M-theory solutions, by considering the strongly coupled limit of
the heterotic string\cite{fh}.  In this case, the role of the ``fifth"
dimension is played by the circle on which the eleventh dimension is
compactified.  They noted that one can choose the chirality of the fermions on
the two walls to be opposite, breaking all of the supersymmetry.  One can then
write down a Witten solution.  Such a configuration is again non-singular.
One can easily write down
solutions when other dimensions
are compactified.  It is not clear, a priori, under what
circumstances these configurations
correspond to approximate solutions of some set
of exact classical equations.  For example, if one compactifies on a
Calabi-Yau space, higher derivative operators will presumably destabilize the
solution (even ``classically").  But, in the spirit of our earlier discussion,
the whole question of the bounce is not particularly interesting unless the moduli are somehow
stabilized, in which case our considerations below will suggest that solutions
exist.  Fabinger and Horava speculated that solutions would exist if moduli
were stabilized.  This view was also espoused by De Alwis and Flournoy\cite{dealwis}.


Returning to our weak coupling compactifications, one might conjecture that
there is some sort of duality between the strongly coupled and weakly coupled
non-supersymmetric strings, and that the BON solutions somehow make sense.
Establishing such a duality is, of course, beyond present techniques.  Indeed,
if one could establish such dualities, one would have presumably established
the non-perturbative consistency of these string states.

So in the weakly coupled limit,it is easy to establish
that a broad class of flat, four dimensional configurations
solve the string equations of motion,
but it is difficult to establish whether a Witten
instability exists. 
In the strongly coupled limit, it is more difficult to determine when
configurations are solutions of the equations of motion, but it is easier to
argue that if they are, there is a non-singular Witten-type solution.  But in
light of these observations it would be surprising if these instabilities were
not generic in non-supersymmetric string solutions.

\section{Including a Potential}

\begin{figure}[t]
\centerline{\epsfxsize=4 in \epsfbox{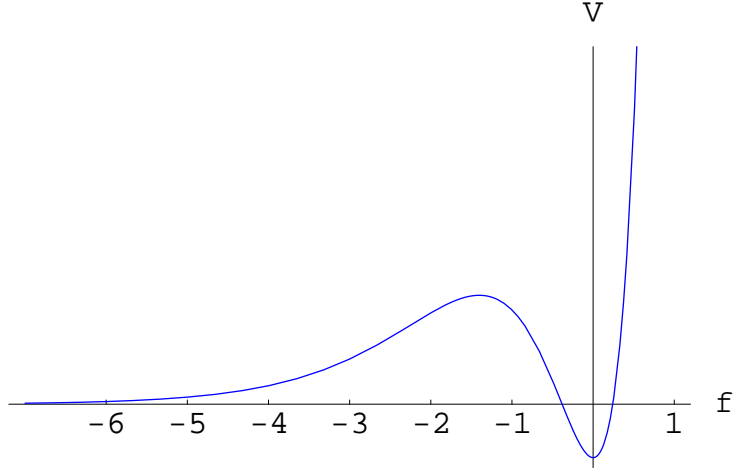}}
\caption{Example potential, $V(\phi)$.}\label{fig:potential}
\end{figure}

In non-supersymmetric theories, a potential for the moduli appears already
at one loop.  This potential tends to zero as $R\rightarrow \infty$, the limit
in which the theory becomes ten (or eleven) dimensional).  It also tends to
zero for small $R$.  This is a reflection of the fact that in string/M theory,
small radius is generally equivalent to large radius, possibly in some
other string theory.  If a non-supersymmetric solution
of string theory is relevant to nature,
the
potential should have at least a stationary point.
In this section we will consider model potentials, with a minimum
of the energy at some value of $R$, and which tends to zero as
$R\rightarrow \infty$ or $R\rightarrow 0$.  An example of a potential
is given in Fig.\ref{fig:potential} (the figure is drawn so as
to emphasize the stationary point, so the behavior at the origin
is not visible).  The Witten solution includes a region
where the extra dimension shrinks to zero size, so the vanishing of the potential
at small radius will be an important feature of the model.\footnote{One can write
down a $T$-dual of Witten's solution, in which the extra dimension blows up as $r \rightarrow
R$.  The existence of this solution raises some conceptual questions which will be
dealt with elsewhere.}

We again want to solve
equations \ref{eqn:cd3},\ref{eqn:cd4}.  We need to think about the
boundary conditions carefully.  It is helpful to return, first,
to five dimensions, and to study a metric of the form:
\beq 
ds^2 = f(r) dr^2 + r^2 d \Omega_3^2 + h(r) d\phi^2.
\eeq
It is straightforward to compute the curvature for this metric.  Vanishing of
the Ricci tensor gives the equations:
\beq
{f^\prime \over f} = -{h^\prime \over h} \\
f^\prime -{2 \over r} f + {2 \over r}f^2=0.
\label{asymptotics}
\eeq 
Assume that $f$ has a singularity (a simple pole) at $r_o$,
\beq
f=a{r_o
  \over 2( r-r_o)}+ {\rm ~constant}, h(r)= 2 r_o (r-r_o) c.
  \eeq
Then near this
point, we have:
\beq
-a + a^2 =0 ~~~{\rm so}~~  a= 1.
\eeq
$h(r)$ is determined by the requirement that the singularity is merely a
coordinate singularity: This gives $c=1$.  Note that at time zero, this is a
configuration of the type considered by Brill and Horowitz.

As boundary conditions at $\xi
\rightarrow 0$, we require that $\phi$ and its first derivative
behave as in the Witten solution.  This insures that, even though
the field and its derivative are blowing up, interpreted
in five dimensions, the singularity is a coordinate singularity.
We then ask whether we can find a solution which asymptotically
tends to the stable vacuum (which we will loosely refer to as the
``false vacuum").  We require that the metric behave as in
Eqn.(\ref{asymptotics}), and treat $r_o$ as a parameter.

\begin{figure}[b]
\centerline{\epsfxsize=2.5 in \epsfbox{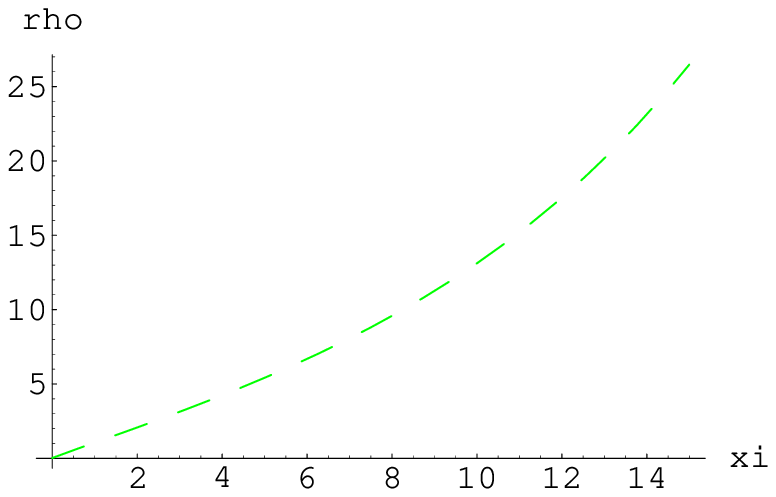} \hskip 0.5in 
\epsfxsize=2.5 in \epsfbox{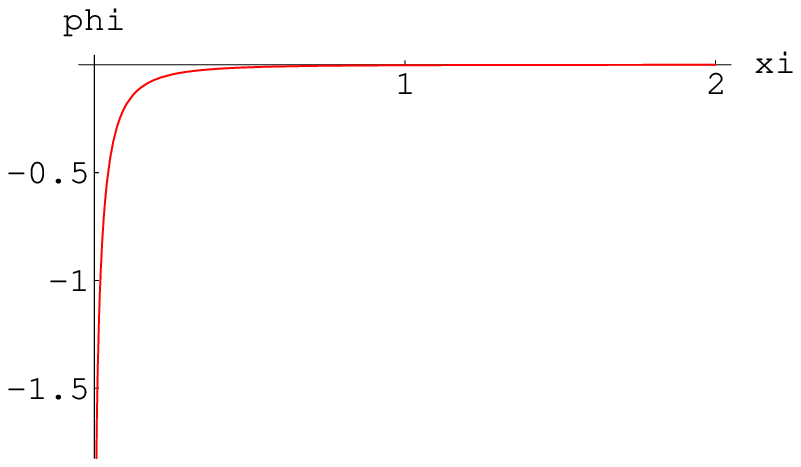}}
\caption{Numerical solution for the Witten bubble with a potential for
the radion.  The dotted line is $\rho(\xi)$ and the solid line is
$\phi(\xi)$.}\label{fig:solution}
\end{figure}

We study the bounce
for a potential $V(\phi)$.  We find a solution, following
Coleman's dictum, by varying $r_o$ so as to obtain both undershoot and
overshoot of $\phi=0$.  We check that the metric is non-singular as
$\rho \rightarrow \infty$.  We find, for the potential of
Fig \ref{fig:potential}
there is a solution.  An example is shown in Fig.\ref{fig:solution}.
We find solutions when the ``false vacuum" has positive, zero, or {\it
negative} cosmological constant.  This is, perhaps, not surprising,
given the existence of configurations of arbitrarily negative energy
in the flat Kaluza-Klein theory.

The existence of an AdS solution is particularly curious.  Since light can
reach the boundary of AdS in finite time, it is natural to ask whether the
wall reaches the boundary in finite time.  As we will explain shortly, the
answer is yes.  From the
perspective of the AdS-CFT correspondence, the existence of such solutions is
also puzzling.  We will discuss these questions further in the concluding
section.

Similar issues have been discussed in \cite{bala,brim}. These authors
cosidered the double analytic continuation of the Schwarzschild black hole in
$AdS_5$ spacetime and calculated the boundary stress tensor. Though their
approach is different from ours some of the concusions they reached overlap
with the present paper.

\subsection{Geodesics in AdS case}

Following a similar analysis to the flat case we can reduce the discussion of
geodesics in the AdS Witten bubble to an analysis of the classical motion of a
particle in a potential.  After solving the CDL problem in 4 dimensions one
can lift the metric back up to 5 dimensions, 
\beq
ds^2=-\rho^2(\xi)dt^2+d\xi^2+\rho^2(\xi) \sinh(t)d\Omega_2^2 +R^2(\xi)dy^2.
\eeq
$\rho$ and $R$ are found numerically for a given potential for the radion,
$\sigma$.  Recall, for instance, that $R=e^{\sqrt{2/3}\sigma}$.  

Using the first 2 transformations of Eq.~(\ref{eqn:transforms}) we can get the
metric into the same form as Eq.~(\ref{eqn:metricprime}), again making the
assumption of fixing one of the angles on the sphere.  We can identify
conserved momenta and find the `conserved energy' of the motion.  We find,
\beq 
\dot{\xi}^2+\frac{k_u k_v}{\rho^2}+\frac{k_y^2}{R^2}=\mu^2
\eeq
with conserved momenta,
\beq
\rho^2\dot{u}=k_v,\ \ \rho^2\dot{v}=k_u,\ \ R^2\dot{y}=k_y.
\eeq
Again one finds that a massive particle is struck by the wall.
This occurs in a finite proper time, no matter how ``close" the particle
starts out to the boundary.  The particle is then bound to the wall.
It takes infinite proper time for the particle to reach the boundary,
but, as we will explain in the next section, only a finite conformal time
elapses until the wall reaches the boundary.

\subsection{Action of the Bubble}

The tunnelling amplitude for the bubble decay is proportional to $e^{-S}$
where $S$ is the classical action of the bounce solution.  For the Witten
bubble this is finite and $S=\frac{\pi R^2}{4G_N}$.  The contribution to this
action from the usual bulk term is zero, because the solution is Ricci flat.
However, a boundary term must be added to cancel second derivative terms this
gives a non-zero contribution.  In general the complete action is,
\beq
S=\frac{1}{32\pi^2 G_N R}\int d^5 x \sqrt{g}R + \frac{1}{16\pi^2 G_N R}\int
d^4 x \sqrt{\gamma}K
\eeq
$\gamma$ is the metric on the boundary with unit normal $\hat{n}$ and 
$K=-\gamma^{ab}\nabla_a \hat{n}_b$ is its intrinsic curvature.

In going from the 5D setup to the 4D gravity-scalar system one must
dimensionally reduce and Weyl rescale the metric.  This results in additional
contributions to both the boundary and bulk actions from the scalar.  Thus,
\begin{eqnarray}
S_4&=&\frac{1}{32\pi^2 G_N R}\int d^4 x \sqrt{\tilde{g}}\left( \tilde{R} +
  \sqrt{\frac{2}{3}} \tilde{\nabla}^2  \sigma - \tilde{g}^{ab}\partial_a
  \sigma\partial_b \sigma +  V(\sigma)\right)\\ &&+ 
\frac{1}{16\pi^2 G_N R} \int d^3 x \sqrt{\tilde{\gamma}}\left( \tilde{K}+
  \frac{1}{2}\sqrt{\frac{2}{3}}\partial_r  \sigma\sqrt{\tilde{g}^{rr}}\right)
\end{eqnarray}

One can check that, in the case where explicit results are known, this gives
the same result for the Witten bubble as the 5D calculation.  In the case of a
non-zero potential, in flat space, the calculation must be done numerically.
We have checked that this gives a finite result for the action.  Likewise, the
action for the 
AdS case must be calculated numerically but also produces a finite result.

\subsection{Spontaneously Broken Supersymmetry}

Now consider supersymmetric string models.  Here one expects that a potential
appears non-perturbatively.  By itself, this is not a qualitative difference
from non-supersymmetric theories.  But if supersymmetry is broken
at low energies (as in gluino condensation and similar mechanisms), the effect
is small, and is negligible for the non-zero Kaluza-Klein modes.
The distinction between supersymmetric and non-supersymmetric solutions at the
classical level has to do with properties of the non-zero modes, so under these
conditions we might expect that, even when supersymmetry is broken, there will
be no BON.

\section{Implications and Conjectures}

One of the most urgent questions in string theory is whether or not low energy
supersymmetry is a prediction.  There are a number of obstructions to deciding
this question.  The first of these is distinguishing approximate
supersymmetric states in string theory from those which are not supersymmetric
in any approximation, or more precisely those which do not have only four
supersymmetries in any limit of an (approximate) moduli space.  The Witten
bubble makes precisely such a distinction.  Our studies, building on earlier
observations of Fabinger and Horava and De Alwis and Flourney, suggest that BON decays are truly
solutions of string theory, and are generic to non-supersymmetric strings.  We
have also seen here that if we {\it assume} that moduli are stabilized, with
vanishing cosmological constant, then typically decays occur.  Vacua for which
this is not the case are presumably not interesting for describing the real
world.  Finally, we have argued that even in the presence of small,
spontaneous supersymmetry breaking, there is no BON instability in the
supersymmetric case.  If these conjectures are correct, then non-supersymmetric
vacua with stable moduli and small cosmological constant would seem an unlikely
outcome of string theory.

Our observations, however, suffer from serious limitations, and must,
for the most part, be viewed as conjectural.  Some can surely be sharpened,
and this is the subject of continuing investigations.  Carefully defining the
appropriate conformal field theories appears possible.  Some of the questions
of decays in supersymmetric and non-supersymmetric states can be studied
in flux compactifications.

However, whether these solutions are indicative of a real pathology or not is
not clear.  First, as always in considering quantum gravity, there is no sharp
argument that it is necessary to include these topologically non-trivial
configurations in some underlying path integral formulation of the theory.
Perhaps they can simply be banished.  We view the existence of a sensible
Euclidean conformal field theory as some evidence that these configurations
are to be included, but this is hardly conclusive.  Second, it could be that,
as pathological as they appear, these pathologies are ultimately smoothed out
in some underlying quantum theory.  All that might be important for our
universe is that our current state be very long lived.

\subsection{AdS Puzzles}

These puzzles and their possible resolutions
 are illustrated by the AdS solutions.  These have a
number of troubling features.  First, the wall reaches the boundary in
finite \emph{conformal} time.  To see this, consider a point on the
leading edge of the wall.  In the CDL coordinates, this can be a point
of large, fixed $\xi$.  For very large
$\xi$, the CDL equations become the equations
for AdS space in somewhat
unusual coordinates.  The transformation between the CDL and global
AdS coordinates, for large $\xi$ and $\psi$ is: 
\beq 
\rho \approx \xi + \psi -\ln(2) +(e^{-2\psi}-e^{-2\xi}) ~~~~~\cos(\tau) \approx 2 e^{-\psi}(1-e^{-2\psi}+2e^{-2\xi}).  
\eeq 
So we see explicitly that we
reach $\rho=\infty$ at time $\tau=\pi/2$.  Curiously, any massive
observer is swept along by the wall, and only reaches the wall after
an infinite \emph{proper} time.  Moreover, if this observer emits
light rays, at large time they take a fixed (proper) time to reach the
boundary.  These observations suggest that it is not be appropriate to
think of the bubble as a bubble in a true AdS space.  One might,
instead, think of starting with the time-symmetric solution (i.e. take
the continuation of the Euclidean solution, and allow $-\infty < \psi
< \infty$), in which, in the far past, one starts, essentially, with
nothing, and returns to that state in the far
future\cite{evaetal}\footnote{We thank Tom Banks for raising this
possibility.}.  In such a space-time, one can still form bubbles (this
is true also for the Minkowski and dS cases), and these will grow,
collide, etc.  Whether there is a sensible quantum theory in such
spaces is not clear.  It seems quite possible that there is not.



There seem to us to be at least three possible resolutions.

\begin{itemize}

\item Stable AdS spaces are expected to have conformal
duals.  If such a dual for these candidate vacua exist,
then there is a stable AdS state, and the solution
we have found can not be interpreted as
a gravitational instability of 
AdS.    Indeed,
since the bubble geometry is not asymptotically ADS, its quantum
description--if one exists--is completely autonomous and we should not
include the BON as a contribution to the gravitational
path integral.  
\item  This leaves the possibility
described above, that the that the bubble of
nothing is a finite-lived quantum system that contracts from nothing and
then expands. Local geometry during the evolution outside of the
bubble looks like ADS space. 
Perhaps there would be some relic of this state in the dynamics of the CFT.
 \item There might be no
quantum completion of the classical solution. In this case the BON is
neither an instability of ADS nor a sensible system in its own right;
it is just an artifact of naive application of the semiclassical method.
\item There is no CFT dual, and there are simply no sensible quantum systems
in cases where a BON solution exists.
\end{itemize} 
One can clearly add further speculations to this list.  For the flat space
and dS case, there are analogous possibilities.

\subsection{Future Directions}

There are other directions for further work, which we are pursuing.
Flux vacua are another arena in which one might expect BON's may
arise, and it is possible that in such vacua one could study decays
with stabilized moduli in a controlled approximation.  One might guess
that the existence of such solutions would depend on whether the
theory, in the limit that certain moduli become large, is or is not
supersymmetric.  Such an analysis will also force
us to sharpen the distinction between
supersymmetric and non-supersymmetric
strings.  Our work also suggests that such solutions will exist
in the case of Randall-Sundrum\cite{RS1,RS2} type models of warped
dimensions.

In light of these observations, reasonable to suggest that
broad classes of non-supersymmetric vacua do not make sense.  This
might be the basis of a prediction of supersymmetry in string theory.
However we have noted alternative interpretations of these solutions
which some readers may find quite plausible.  These issues seem
sufficiently intriguing -- and potentially important -- to be worthy
of further study.



\noindent
{\bf Acknowledgements:}

\noindent
We thank Anthony Aguirre, Tom Banks, Raphael Bousso, Willy Fischler, Gary
Horowitz, Scott Thomas, and Ed Witten for discussions.  This work supported in
part by the U.S.  Department of Energy.


\end{document}